\documentclass{aip-cp}
\usepackage{graphicx}
\usepackage[numbers]{natbib}
\usepackage{rotating}
\usepackage{color}
\begin{document}
%
\title{First Results from The GlueX Experiment}
\author{The GlueX Collaboration}
\author[Florida State University]{H.~Al Ghoul}
\author[Athens University of]{E.~G.~Anassontzis}
\author[Jefferson Lab]{F.~Barbosa}
\author[Connecticut University of]{A.~Barnes}
\author[Regina University of]{T.~D.~Beattie}
\author[Indiana University]{D.~W.~Bennett}
\author[MEPhI]{V.~V.~Berdnikov}
\author[North Carolina Wilmington University of]{T.~Black}
\author[Florida International University]{W.~Boeglin}
% ...
\affil[Arizona State University]{Arizona State University, Tempe, Arizona 85287, USA}
\affil[Athens University of]{National and Kapodestrian University of Athens, 15771 Athens, Greece}
\affil[Carnegie Mellon University]{Carnegie Mellon University, Pittsburgh, Pennsylvania 15213, USA}
\affil[Catholic University of America]{Catholic University of America, Washington, D.C. 20064, USA}
\affil[Connecticut University of]{University of Connecticut, Storrs, Connecticut 06269, USA}
\affil[Florida International University]{Florida International University, Miami, Florida 33199, USA}
\affil[Florida State University]{Florida State University, Tallahassee, Florida 32306, USA}
\affil[Glasgow University of]{University of Glasgow, Glasgow G12 8QQ, United Kingdom}
% ...
\affil[Indiana University]{Indiana University, Bloomington, Indiana 47405, USA}
\affil[Jefferson Lab]{Thomas Jefferson National Accelerator Facility, Newport News, Virginia 23606, USA}
\affil[MEPhI]{National Research Nuclear University Moscow Energineering Physics Insitute, Moscow 115409, Russia}
\affil[Massachusetts Amherst University of]{University of Massachusetts, Amherst, Massachusetts 01003, USA}
\affil[Massachusetts Institute of Technology]{Massachusetts Institute of Technology, Cambridge, Massachusetts 02139, USA}
\affil[Norfolk State University]{Norfolk State University, Norfolk, Virginia 23504, USA}
\affil[North Carolina AT State]{North Carolina A\&T State University, Greensboro, North Carolina 27411, USA}
\affil[North Carolina Wilmington University of]{University of North Carolina at Wilmington, Wilmington, North Carolina 28403, USA}
\affil[Northwestern University]{Northwestern University, Evanston, Illinois 60208, USA}
\affil[Regina University of]{University of Regina, Regina, Saskatchewan, Canada S4S 0A2}
\affil[Santa Maria University]{Universidad T\'ecnica Federico Santa Mar\'ia, Casilla 110-V Valpara\'iso, Chile}
% ...
\corresp[cor1]{Corresponding author: cmeyer@cmu.edu}

\maketitle

\begin{abstract}
The GlueX experiment at Jefferson Lab ran with its first commissioning beam in late 2014
and the spring of 2015. Data were collected on both plastic and liquid hydrogen targets, and
much of the detector has been commissioned. All of the detector systems are now performing
at or near design specifications and events are being fully reconstructed, including exclusive 
production of $\pi^{0}$, $\eta$ and $\omega$ mesons. Linearly-polarized photons were successfully 
produced through coherent bremsstrahlung and polarization transfer to the $\rho$ has been observed.
\end{abstract}

\section{INTRODUCTION}
The GlueX Experiment~\cite{gluex-ref} is a key element of the Jefferson Lab 12 GeV upgrade. The 
experiment is at the end of a new beamline from the Continuous Electron Beam Accelerator Facility
(CEBAF) at Jefferson Lab, that will use 12 GeV electrons to deliver linearly-polarized photons to a new 
experimental area, Hall D. The primary physics goal of GlueX is to discover and study the properties 
of hybrid mesons--particles where the gluonic field contributes directly to the $J^{PC}$ quantum 
numbers of the mesons~\cite{meyer:2015eta}. Lattice QCD calculations indicate that several of the 
nonets of these hybrid mesons have exotic quantum numbers, forbidden $J^{PC}$ for a simple 
fermion-antifermion system~\cite{Dudek:2013yja}. In addition, the expected masses for the lightest 
hybrids are well-matched to the energy and kinematics accessible to 
the GlueX experiment. Commissioning of the GlueX experiment started in late 2014 and continued
through the spring of 2015.  The first physics quality beam is expected in 2016. In this paper we 
report on the performance achieved both in calibrating and understanding the detector 
and present the measurement of the transfer of the photon's polarization to photo-produced $\rho$ 
mesons. The results from a few hours of running with polarized photons are consistent with older 
measurements~\cite{Criegee:1969qg,Ballam:1970qn} and comparable in statistics to the older data.

\section{THE GLUEX EXPERIMENT}
\begin{figure}[h!]
  \centerline{\includegraphics[width=300pt]{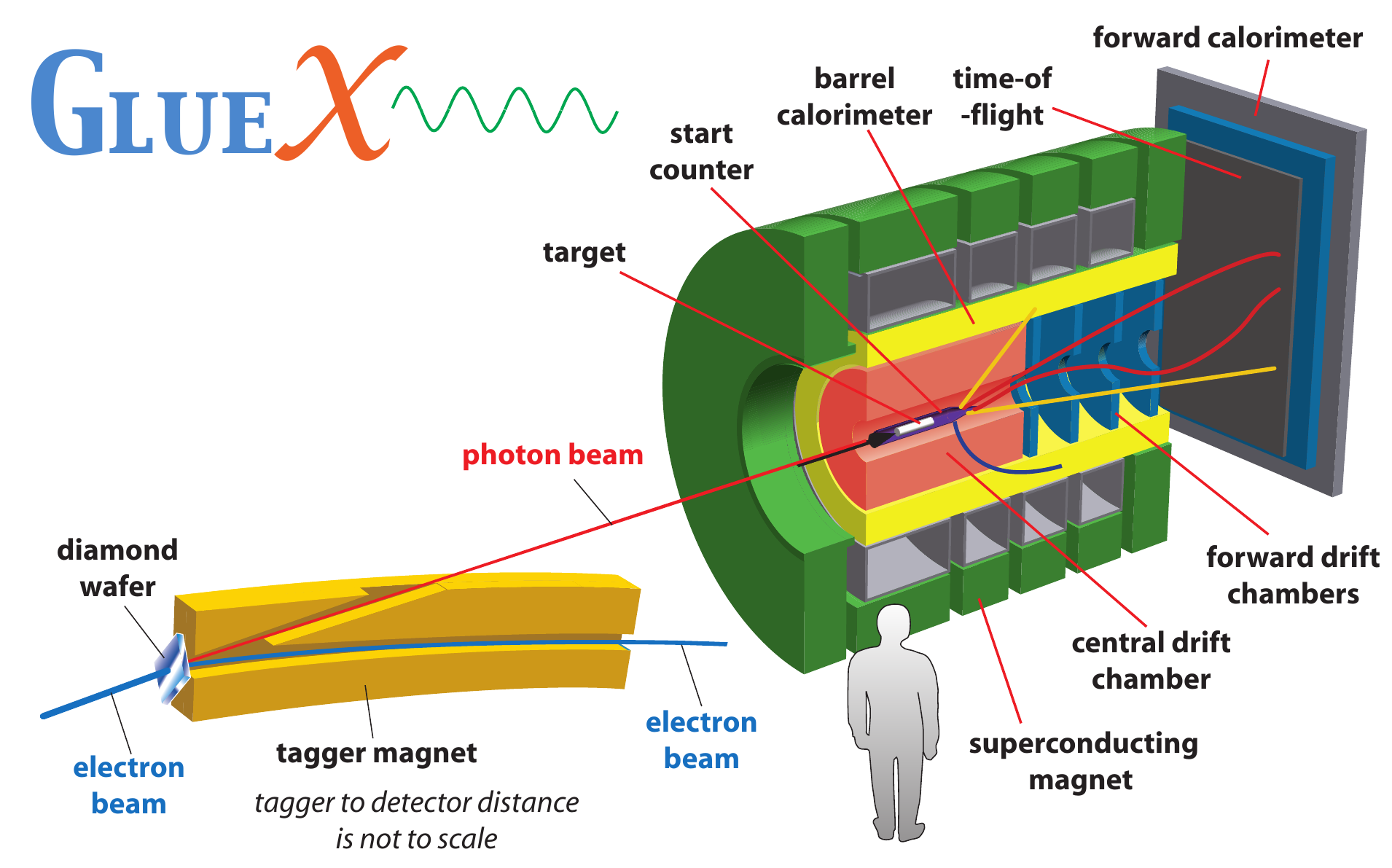}}
  \caption{\label{fig:detector}A schematic drawing of the Hall-D photon tagger and the GlueX detector
at Jefferson Lab. See text for more information.}
\end{figure}

\noindent\paragraph{The Hall D Photon Beam}
The new Hall D facility has been designed as a photon-only hall, taking advantage of the $12$~GeV 
electron beam from the upgraded CEBAF. The highest-energy electrons in the CEBAF accelerator 
are extracted and transported to the Hall D tagger hall. Here, they pass through a bremsstrahlung 
radiator, and then into a dipole magnet for tagging the energy of the scattered electrons. Those
electrons not interacting in the bremsstrahlung radiator are deflected into an electron dump attached
to the tagger hall. For the electrons producing a bremsstrahlung photon in the 
region between $25$\% and $98$\% of the primary electron beam energy, the scattered electrons 
are detected in pair of tagger hodoscopes, thus tagging the energy of the photon. For $12$~GeV electrons,
 the fine-grained \emph{tagger microscope} (TAGM) has 
been placed to tag photons in the $8.2$~GeV to $9.2$~GeV range, with each microscope element
spanning $10$~MeV in energy. For both higher- and lower-energy photons, the \emph{tagger 
hodoscope} (TAGH) tags photons using elements about $30$~MeV wide. This is shown in the 
left-hand side of Figure~\ref{fig:detector}.

Several different bremsstrahlung radiators are utilized in GlueX. The first are thin pieces of 
aluminum ($1.5\, \mu m$, $10\, \mu m$ and $30\, \mu m$), while the second is a $20\, \mu m$ 
thick diamond\footnote{During the 
spring 2015 running of GlueX at lower electron energies, $50\,\mu m$ and $100\,\mu m$ thick 
diamonds were utilized.} crystal. The diamond is aligned to produce coherent bremsstrahlung in
the $8.4$~GeV to $9.0$~GeV range. These coherent photons are linearly polarized relative to the
crystal axes in the diamond. The peak polarization is expected to be $40\%$, while the average 
polarization for $8.4$ to $9.0$~GeV photons is expected to be $36\%$. The diamond can be
rotated to produce two perpendicular directions of linearly-polarized photons. The 
aluminum radiator produces a standard bremsstrahlung spectrum with the 
characteristic $1/E_{\gamma}$ fall off. Coherent bremsstrahlung photons are produced in the 
direction of the incident electrons, while incoherent bremsstrahlung photons are produced in a cone 
around that direction.

The photons travel through an $80$~m long vacuum beamline before entering Hall D. 
Here, the off-axis photons are removed by passing the beam through a $3.4$~mm diameter 
collimator\footnote{During the spring 2015 run, a $5$~mm diameter collimator was used.}
that has been instrumented around the hole with an \emph{active collimator}~\cite{Miller_1973}
that provides fast feedback on the beam positioning. Photons passing through a primary collimator
a secondary collimator, and then pass through a \emph{triplet polarimeter} that is used to monitor 
the linear polarization. The photons then travel into a \emph{pair spectrometer} 
system~\cite{pair-spect} that is used to monitor both the energy and intensity of the photon beam. 
The photons then travel into the GlueX detector, with some interacting in the liquid hydrogen physics 
target. The remainder pass into the photon beam dump at the back of Hall D.

Initial operation of the GlueX experiment will run at a beam intensity of $10^{7}$ photons per second in 
the coherent energy peak ($8.2$~GeV to $9.2$~GeV), with the ultimate rate limit expected to be an
order of magnitude higher. Operations during commissioning runs were at lower intensities.
\noindent\paragraph{The GlueX Detector}
The GlueX detector is azimuthally symmetric and nearly hermetic for both charged particles and 
photons, and is shown in the right-hand side of Figure~\ref{fig:detector}. As noted above, the 
photon beam enters and is incident on a $30$~cm long liquid hydrogen target which was successfully
operated with the photon beam in the spring of 2015. The largest element of the GlueX detector is a 
solenoidal magnet, providing a magnetic field of about $2$~T along the direction of the beam. 
Charged particles from the primary interaction first pass through the \emph{start counter}, a 
$30$-element scintillator detector that surrounds the hydrogen target and tapers towards the 
beamline in the forward direction. Light is detected using Hamamatsu Multi Pixel Photon 
Counters~\cite{Barbosa-2012,sipm} (silicon photomultipliers). 
The start counter measures a track timing with 
$300$~ps resolution, which is sufficient to identify the bucket from the electron beam $499$~MHz 
radio frequency (RF) structure from which the primary electron came. The start counter also provides
information for track reconstruction and some particle identification capability through energy 
loss measurements. 

Immediately surrounding the start counter is the \emph{central drift 
chamber} (CDC)~\cite{cdc}. This detector is a $28$ layer device based on straw-tube technology that
includes both axial ($12$) and stereo ($8$ at $6^{\circ}$ clockwise and $8$ at $6^{\circ}$
counter clockwise) layers with a total of $3522$ straws. The CDC provides position measurements
along the charged tracks with $150\, \mu m$ accuracy in the $r-\phi$ plane, and at the $mm$ level 
along the beam axis by utilizing the stereo information. The detector also provides $dE/dx$ information
for charged tracks and has good $\pi/p$ separation up to $p\sim1$ GeV$/c$. The CDC is $150$~cm long 
and has active elements between $10$~cm and $59$~cm radially.

Downstream of the CDC are the four packages of the \emph{forward drift chamber} system 
(FDC)~\cite{fdc}. Each package is based on six layers of planar drift chambers with both anode and cathode 
readouts. The dual readout allows each layer to reconstruct a three-dimensional space point which
is very helpful for tracking in the high-magnetic field environment. The FDC provides position
measurements of charged tracks at the $200\, \mu m$ level as well as providing $dE/dx$ information
for particle identification. The four detector packages are placed so as to allow charged particle track 
reconstruction down to a polar angle of about $1^{\circ}$ from the beamline. Both drift chambers 
systems are read out using custom $125$~MHz Flash ADC systems~\cite{flash125}.  

Surrounding the tracking devices inside the solenoid is the \emph{barrel calorimeter} 
(BCAL)~\cite{bcal-1,bcal-2,bcal-3}. This is a lead-scintillating fiber calorimeter with readout 
on both the upstream and downstream layers. For particles entering normal to the calorimeter 
face, the calorimeter is $14.9$ radiation lengths thick. The BCAL is sensitive to photons between 
polar angles of $12^{\circ}$ and $160^{\circ}$ and $\pi^{0}$ have been reconstructed using $\pi^{0}$ decay 
photons of energy below $100$~MeV. The expected energy resolution is 
$\sigma E/E\approx\, 5.4\%/\sqrt{E} + 2.3\%$ 
and the typical width of reconstructed $\pi^{0}$ for GlueX reactions is about $9$~MeV.  
Due to its proximity to the solenoidal field, the BCAL is read out using the same Hamamatsu 
Multi Pixel Photon Counters as in the start counter~\cite{sipm}. The BCAL is read out using both $250$~MHz custom Flash 
ADCs as well as $60\, ps$ TDCs. The timing information is used to provide time-of-flight
information for particles interacting in the BCAL, and is used in the global particle identification
methods for both charged particles and photons.

The second calorimeter system in GlueX, the \emph{forward calorimeter} (FCAL), is located 
downstream of the solenoidal magnet~\cite{fcal-1}. This consists of $2800$ lead-glass crystals, 
each $45$~cm long ($14.5$ radiation lengths). The crystals are stacked such that the active area 
of the FCAL is circular, and spans polar angles from $12^{\circ}$ down to $1^{\circ}$. The FCAL is 
sensitive to photons of $E_{\gamma}\geq 100$~MeV, which is well-matched to the kinematics 
of GlueX. The energy resolution of the FCAL is expected to be 
$\sigma_{E}/E\approx 5.6\%/\sqrt{E}  + 3.5\%$.

Upstream of the FCAL is the \emph{time-of-flight wall} (TOF), which consists of two layers of
$2.5$~cm thick scintillator bars with readout using photomultiplier tubes on both ends of the detector. 
Signals from the TOF are processed in both high-resolution TDCs and Flash ADCs to provide both
timing and energy-loss information. The timing resolution of the TOF system is $100$~ps, providing 
particle identification for polar angles from $12^{\circ}$ down to $1^{\circ}$.

\subsection{DETECTOR PERFORMANCE}
During the GlueX commissioning runs in late 2014 and spring 2015, nearly two billion triggered
events were recorded under a variety of trigger, detector, and magnetic field configurations.
In addition, triggering on cosmic events has been ongoing since the summer of 2014. All of these data
have been used to align and calibrate the detector, as well as assess the overall performance of
the detector. 
\paragraph{The Photon Beam}
The TAGM and TAGH systems are used to determine the energy of the beam photon. The combined
system is checked using the pair spectrometer, which uses a convertor to produce an $e^{+}e^{-}$
pair from a primary photon, and then detects the electrons and positrons in coincidence. A special trigger
looks for events in the pair spectrometer and then reads out the pair spectrometer system together
with TAGH and TAGM. This allows us to determine and monitor the energy calibration of the 
tagger system. Energy resolutions of both the TAGH and TAGM systems are given by detector
geometry and alignment and design resolution has been reached.
\begin{figure}[h!]
  \centerline{\includegraphics[width=250pt]{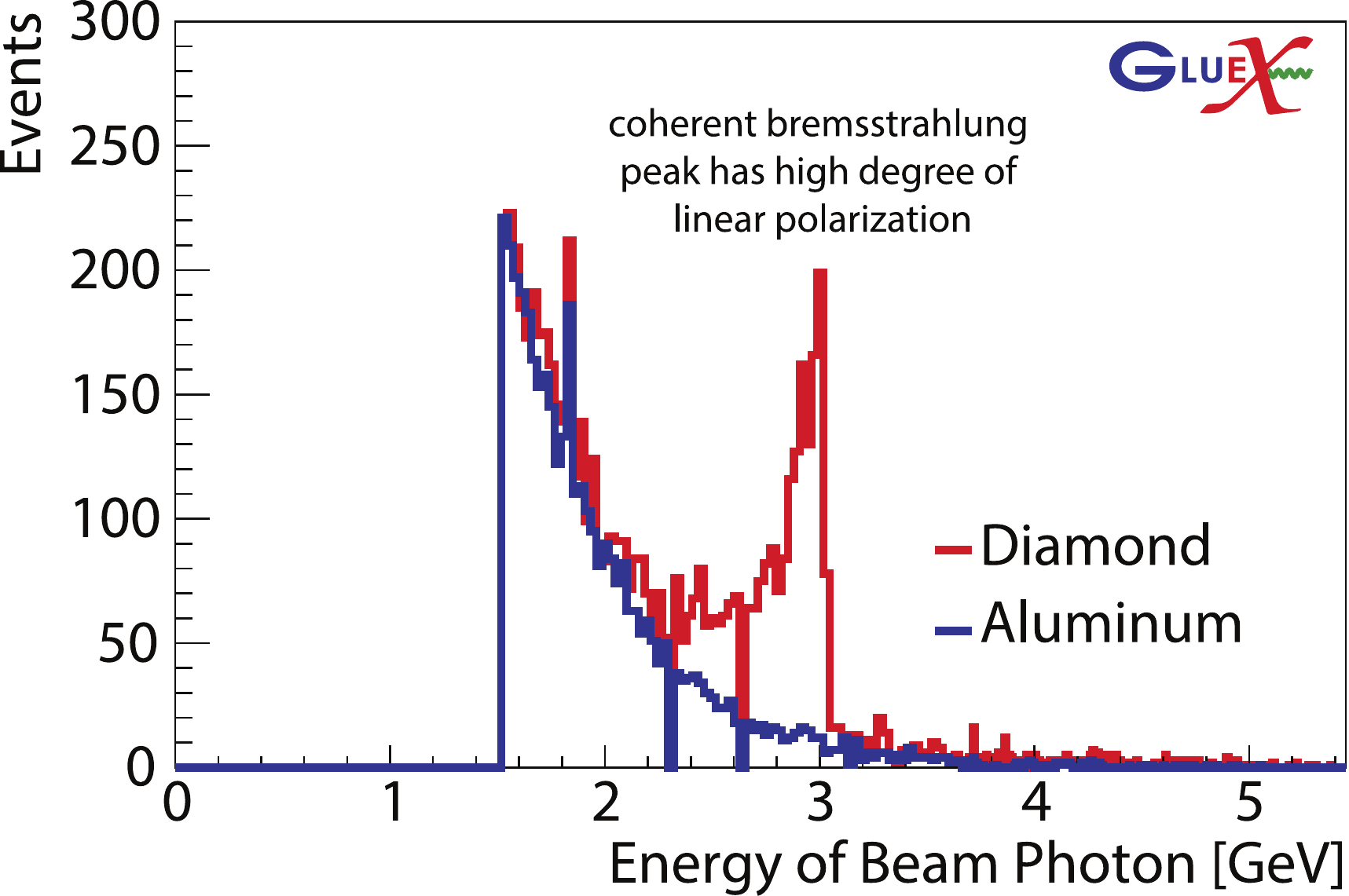}}
  \caption{\label{fig:photon_energy} The number of events as a function of photon beam energy
for the aluminum radiator (blue), and a thin diamond radiator (red). The $1/E_{\gamma}$
intensity fall off is clearly seen for the aluminum radiator, while the coherent enhancement is
clearly seen for the diamond radiator.}
\end{figure}

In addition to the photon energy, the linear polarization of
the photon beam is also important. This aspect of the beam has been commissioned using a $5.5$~GeV
electron beam and a $50\,\mu m$ thick diamond radiator. Figure~\ref{fig:photon_energy} shows
the energy spectrum of photons for an amorphous radiator and the diamond radiator. The expected
$1/E_{\gamma}$ spectrum is seen for the amorphous radiator, while the coherent enhancement at 
about $3$~GeV energy is observed with the aligned diamond radiator. Using the shape of the coherent 
peak, the linear polarization is calculated to be $60$\% at the coherent edge and an average
of $50$\% over photon energies in the $2.5$ to $3.0$~GeV range. An independent 
measurement using the triplet polarimeter yielded a consistent result. Finally, as discussed later,
measurement of the $\rho$ polarization also yielded a photon polarization consistent with the 
other measurements.

\begin{figure}[h!]\centering
\includegraphics[width=0.85\textwidth]{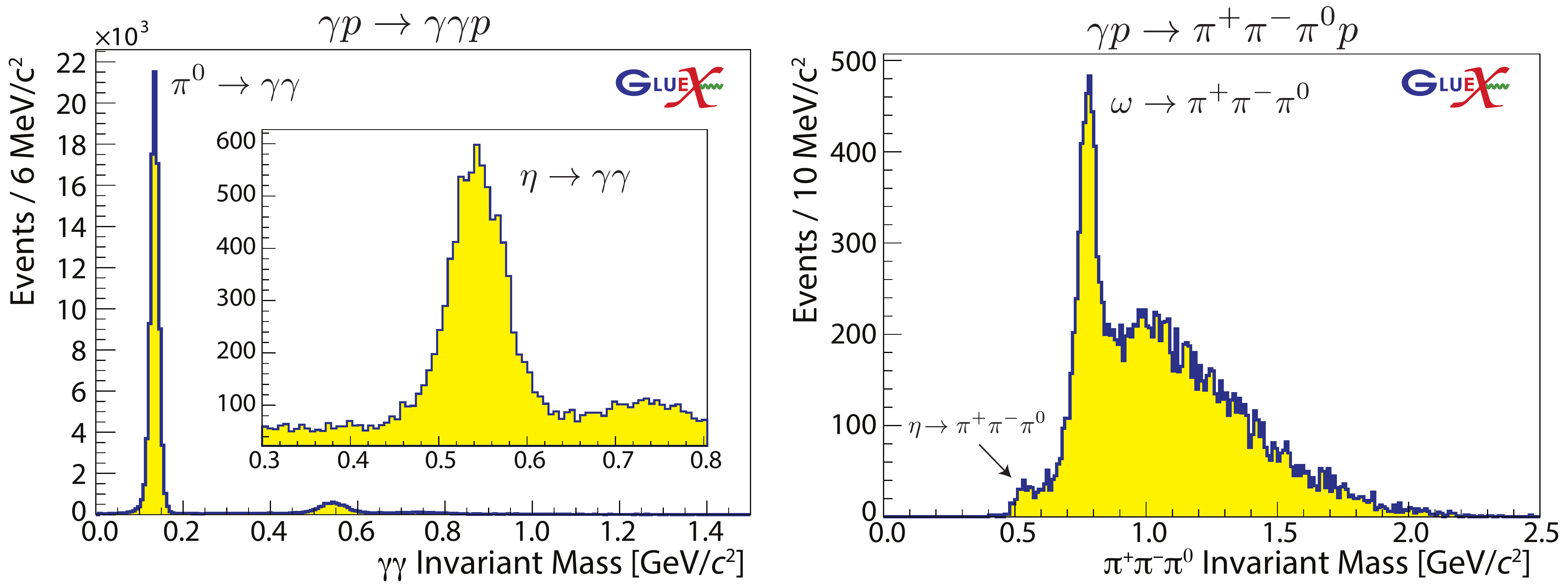}
\caption{\label{fig:pi-0}
(Left) The $\pi^{0}$ and $\eta$ masses as determined from a pair of photons
from $\gamma p \rightarrow p \gamma \gamma$ events. The lower-left image shows 
the full invariant mass range where the $\pi^{0}$ and $\eta$  peaks are visible. The
insert on the upper right zooms in on the $\eta$ peak. (Right) The $\pi^{+}\pi^{-}\pi^{0}$ invariant 
mass from $\gamma p \rightarrow p \pi^{+}\pi^{-}\gamma\gamma$ events where $m_{\gamma\gamma}$ 
is consistent with the mass of the $\pi^{0}$. In both reactions, the detected photons can be in both
the BCAL and the FCAL.}
\end{figure}

\paragraph{The GlueX detector}
Listing detectors in the same order as above, calibration of the start counter has achieved
a resolution better than the $300$~ps design resolution and is sufficient to identify the RF beam
bucket of the primary electron. The CDC has achieved a resolution of $200\, \mu m$ prior
to correcting for straw deformations. Ongoing work on these latter corrections shows that 
the resolution will likely be better than the design goal of $150\,\mu m$. For the FDC, the
cathode resolution has reached the design value of $200\, \mu m$, while the anode resolutions
have reached $150\, \mu m$, exceeding design specifications. For events with multiple tracks
from the same vertex, the vertex positions were reconstructed with a resolution $3$-$4$~mm
averaged over all tracks. For the calorimeters, energy calibration procedures rely on 
optimizing the reconstructed $\pi^{0}$ mass by adjusting appropriate constants. While the
methods have been shown to work, the sizable data set needed to carry out the calibration 
is not yet available. An example of current calorimeter performance is shown in the left-hand
plot in Figure~\ref{fig:pi-0}, where both the $\pi^{0}$ and $\eta$ are reconstructed from
their two photon decays for events from the exclusive final state
$\gamma p \rightarrow p \gamma\gamma$. Currently, a  $\pi^{0}$ width better than $9$~MeV 
has been achieved
using photons detected in the BCAL and in the FCAL. In the right-hand plot is shown the 
$\pi^{+}\pi^{-}\pi^{0}$ invariant mass for events from the exclusive final state
$\gamma p \rightarrow p \pi^{+}\pi^{-}\gamma\gamma$. A clear signal for the 
$\omega\rightarrow \pi^{+}\pi^{-}\pi^{0}$ is observed, where the width of the $\omega$ peak 
is found to be well-described by a Breit-Wigner peak, with the $8.49$~MeV width of the $\omega$ 
convoluted with a Gaussian with a resolution of $\sigma=31.5$~MeV.

\begin{figure}[h!]
\includegraphics[width=0.85\textwidth]{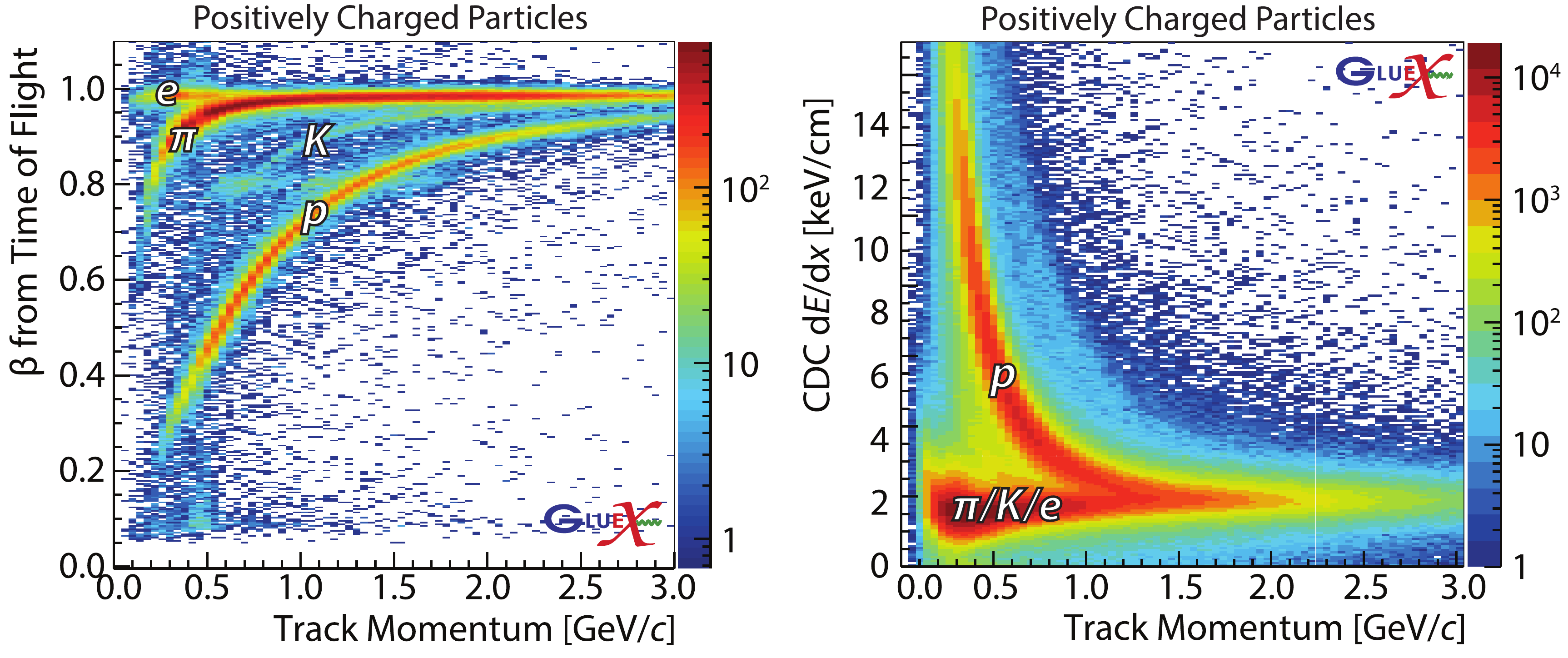} 
  \caption{\label{fig:tof_positive} For positively charged particles: (Left) the $\beta$ as determined from both time-of-flight and
path-length measurements in the detector versus the reconstructed particle momentum and (Right)
energy loss $dE/dx$ in the CDC as a function of reconstructed particle momentum.}
\end{figure}

The TOF system has also achieved $100$~ps timing resolution for charged particles that are
detected in both planes of the system. One of the primary uses of the TOF is to help with 
particle identification. This process involves not only a flight time, as determined from the 
RF beam bunch and the TOF, but also the path length and momentum of the charged 
particle as determined by the CDC and FDC systems. Figure~\ref{fig:tof_positive} shows 
one way to evaluate particle identification, where the calculated $\beta$ of a charged
particle is plotted against its momentum. The figure shows positively charged particles,
and clear bands are seen for $e^{+}$, $\pi^{+}$, $K^{+}$, and protons, with clear proton 
identification up to $3$~GeV$/c$ and $\pi /K$ separation up to about $2$~GeV$/c$. For charged
particles traversing the CDC, energy-loss measurements ($dE/dx$) are made concurrently 
with each position measurement. The right hand image in Figure~\ref{fig:tof_positive} shows the 
measured $dE/dx$ for positively-charged particles as a function the particle's momentum. Clear 
$\pi/p$ separation is evident up to about $1\,$GeV$/c$.

\section{POLARIZATION TRANSFER TO THE $\rho$ MESON}
\begin{figure}[h!]
\begin{tabular}{cc}
\includegraphics[width=0.375\textwidth]{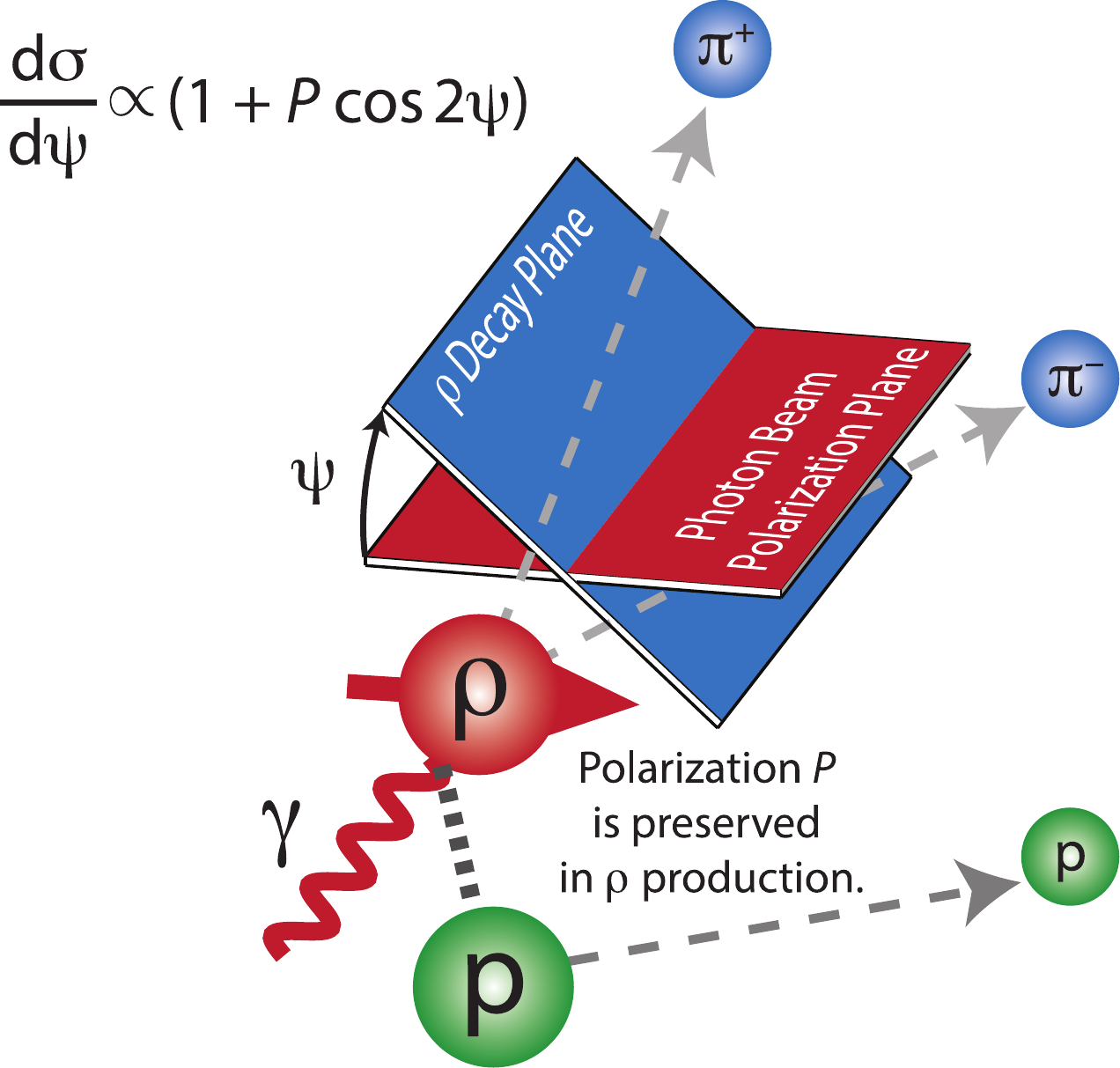}
&
 \includegraphics[width=0.525\textwidth]{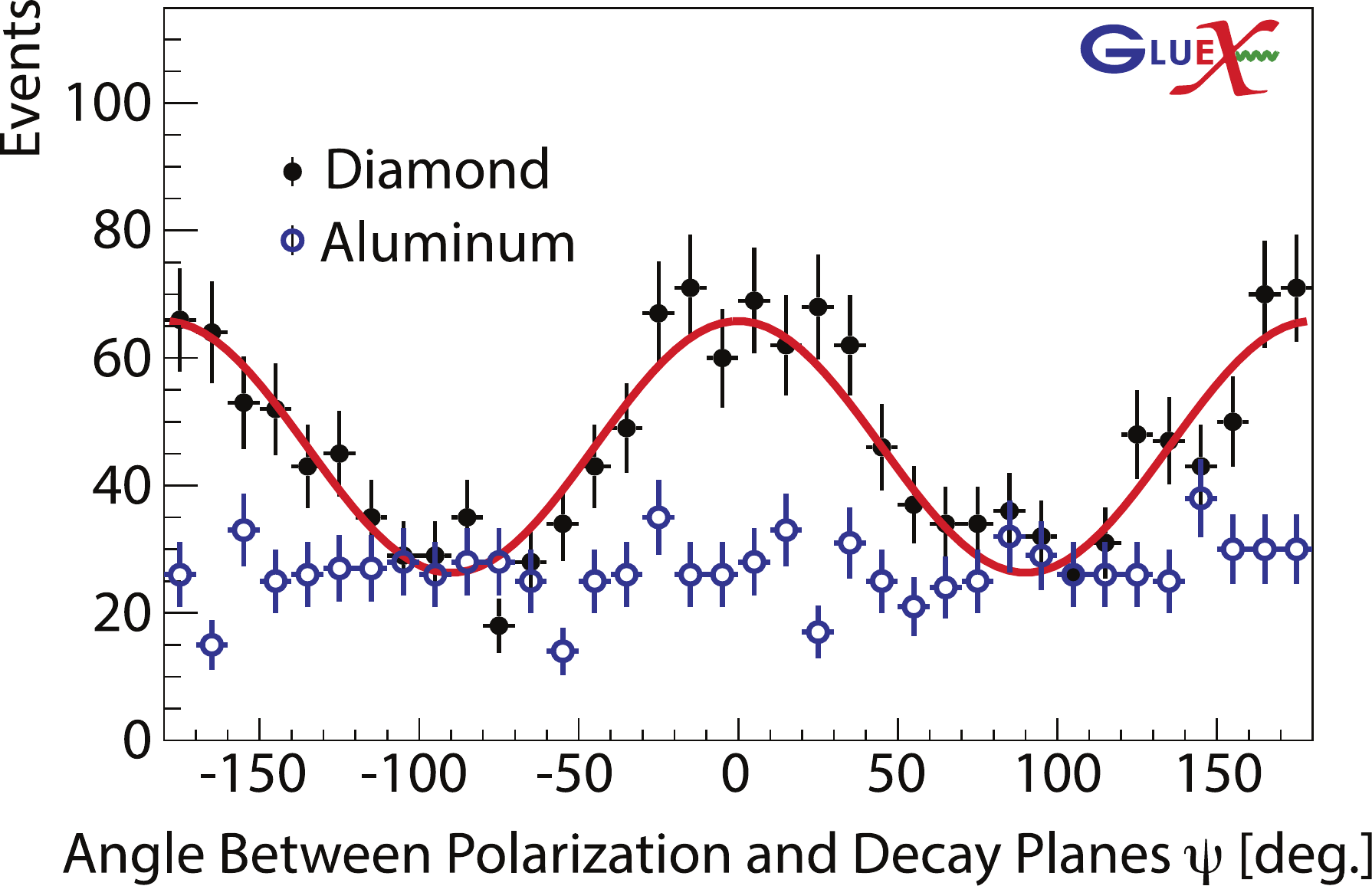}
\end{tabular}
\caption{\label{fig:rho_decay}The decay angular distribution for the $\rho$ meson for 
linearly polarized photons (red curve on solid black markers) and for unpolarized photons (blue open markers).}
\end{figure}

For vector meson photoproduction using linearly polarized photons, one of the observables
that can be measured is the beam asymmetry, $\Sigma$, which can be interpreted as the 
percentage transfer of the linear polarization of the photon to the vector meson.
If the spin-density matrix elements of the vector meson can be measured~\cite{schilling},
then $\Sigma$ can be written directly in terms of them.
\begin{eqnarray*}
\Sigma & = & \frac{\rho^{1}_{11} + \rho^{1}_{1-1}}{\rho^{0}_{11}+\rho^{0}_{1-1}} \, ,
\end{eqnarray*}
The measurement of the $\rho^{1}_{ij}$ requires linearly polarized photons. Alternatively,
the polarization transfer can be approximately determined by measuring the angular 
distribution of the vector meson decay through 
\begin{eqnarray*}
\frac{d\sigma}{d\psi} & \propto & 1 + P\,\Sigma\, \cos2\psi \, ,
\end{eqnarray*}
where $P$ is the linear polarization of the photon beam and $\psi$ is the azimuthal angle between the 
photon polarization vector and the decay plane of the vector meson. This angle is illustrated in the 
left-hand aide of Figure~\ref{fig:rho_decay}. 
For the case of the $\rho$ meson, $\Sigma$ has been measured to be close to unity for the 
photon energies of interest~\cite{Ballam:1970qn,ballam-72,ballam-73}. 

\begin{figure}[h!]
\includegraphics[width=0.375\textwidth]{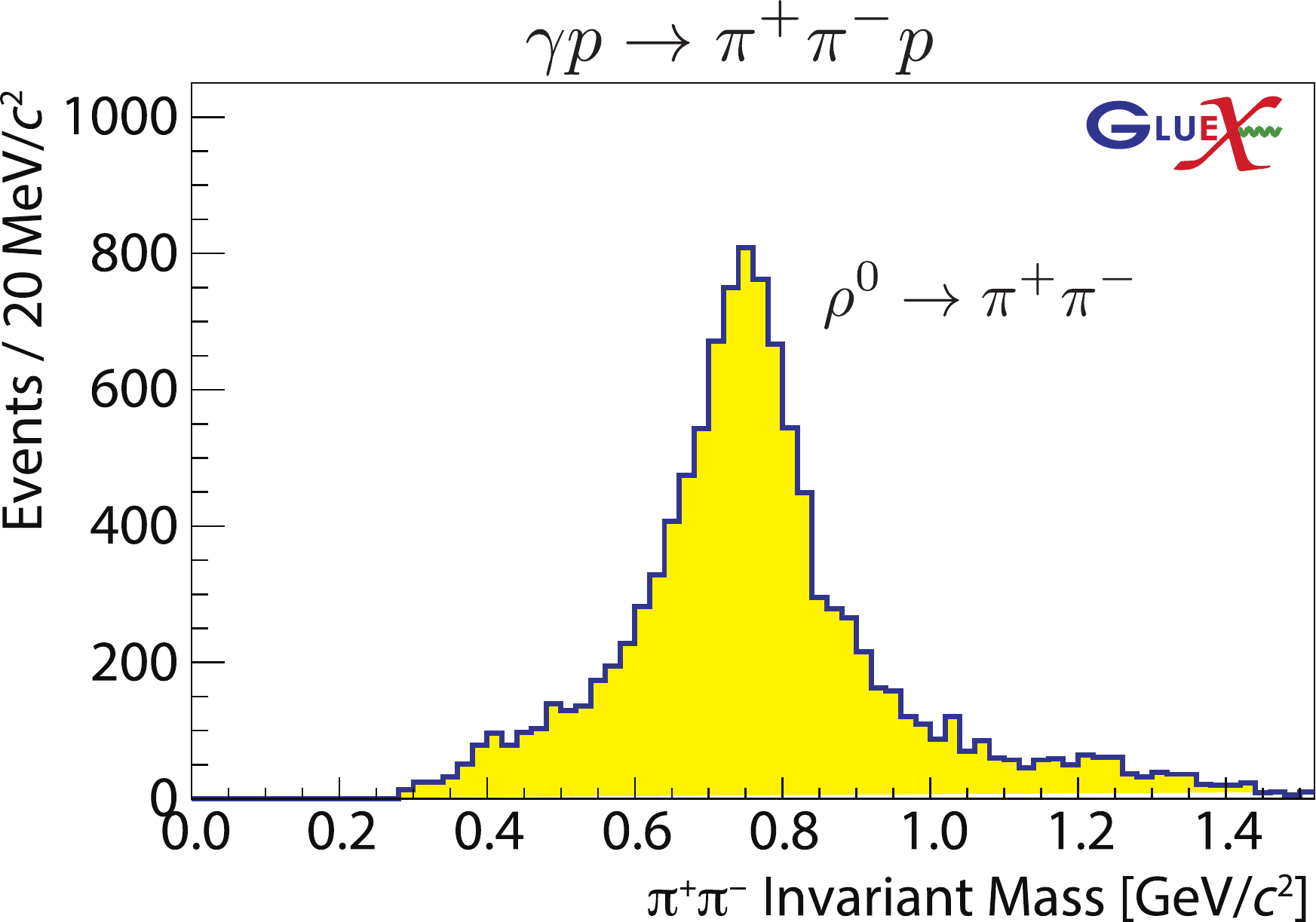}
\caption{\label{fig:pi-plus-pi-minus} The reconstructed $\pi^{+}\pi^{-}$ invariant mass for
events from the exclusive reaction $\gamma p\rightarrow p\pi^{+}\pi^{-}$.}
\end{figure}

Using a $5.5$~GeV electron beam and a $50\, \mu m$ thick diamond radiator, linearly polarized
photons were produced in the $2.5$ to $3.0$~GeV energy range. From the all the collected 
events, we selected exclusive $\gamma p \rightarrow \pi^{+}\pi^{-}p$ events utilizing many of 
the detector systems described in the previous sections.  Charged particle tracks for the proton, 
$\pi^{+}$ and $\pi^{-}$ were reconstructed using the CDC and FDC.  For proton tracks we required 
a large $dE/dx$ measured in the CDC, which provides good separation from $\pi^{+}$ tracks up to 
$p\sim1$ GeV/c (see the right-hand image in Figure~\ref{fig:tof_positive}).  As the tagged, 
initial-state photon energy and the final-state particle momenta are all independently measured, 
the event kinematics are over-constrained.  Thus, momentum conservation was enforced by utilizing 
the magnitude of the difference between the initial and final state 4-momenta, requiring 
$|(p_{final}-p_{initial})^2| < 0.02\, \mathrm{GeV^2/c^4}$.  The resulting $\pi^{+}\pi^{-}$ invariant mass 
distribution is shown in Figure~\ref{fig:pi-plus-pi-minus}. For determining the polarization 
transfer to the 
$\rho$ we selected $\pi^{+}\pi^{-}$ candidates with 
$0.60 < M_{\pi^+\pi^-} < 0.95 \,\mathrm{GeV/c^2}$, and beam 
photons with $2.5 < E_{\gamma} < 3.0$ GeV, where the coherent bremsstrahlung is enhanced 
(see Fig.~\ref{fig:photon_energy}).

The measured decay distribution for both unpolarized and linearly polarized photons is shown in 
the right-hand side of Figure~\ref{fig:rho_decay}. For the case of the aluminum radiator 
(unpolarized photons), the angular distribution is uniform, while for the diamond radiator (linearly 
polarized photons), there is a clear $\cos\, 2\,\psi$ dependence.  These azimuthal decay distributions 
are not corrected for acceptance, however these corrections should be small due to the azimuthal 
symmetry of the GlueX detector, which is nicely demonstrated by the uniform distribution in the 
case of the aluminum radiator.  As noted earlier, taking the value of $\Sigma$ from earlier 
measurements one can determine the linear polarization of the photon beam, $P$, extracted from 
a fit to this angular distribution is consistent with the photon polarization determined 
from the shape of the coherent peak as described earlier. The number of events in 
Reference~\cite{Ballam:1970qn} is similar to what was obtained in GlueX in a few hours of 
low-intensity commissioning running. This observation of the polarization transfer to the $\rho$
provides clear evidence that with full intensity, significant new physics results from GlueX will be 
possible.

\section{SUMMARY}
Commissioning of the Hall D beamline and the GlueX experiment at Jefferson Lab is
well underway.  Calibrations of the detector systems are in an advanced state, with many systems 
at, or exceeding, design specifications. All systems are within $30\,\%$ of design specifications,
and once sufficient data are obtained to complete calibrations, all are expected to reach design 
specifications. While the hardware trigger allowed for substantial data taking, improvements
are still being made. While early commissioning data ran with  a very-large data footprint
to enable studies of the acquisition and data transfers, the experiment will be implementing
the final compact formats in late 2015 and 2016, and rates are expected to reach design
expectations. Offline analysis of data has also rapidly evolved, and a smooth system is now
in place to rapidly process production data and produce a very compact analysis format. These
later data have been used to extract physics from GlueX, as evidenced by the measured 
decay distribution of the $\rho$. 
 
\section{ACKNOWLEDGMENTS}
The authors thank the staff and administration of the Thomas Jefferson National Accelerator
Facility who made this experiment possible. This work was supported in part by the U.S.
Department of Energy (under grants
DE-FG02-87ER40315, 
DE-FG02-92ER40735, 
DE-FG02-94ER40818 , 
DE-SC0010497, 
DE-FG02-87ER40344, 
and DE-FG02-05ER41374); 
the National Science Foundation (under grants 
PHY-1207857, 
PHY-1508238,  
PHY-1306418,  
PHY-1206043,  
PHY-1506303, 
PHY-1507208, 
and PHY-1306737); 
the Natural Sciences and Engineering Research Council of Canada (NSERC) (under grant numbers
SAPJ-326516 and 
SAPIN-105851-2011), 
the Russian Foundation for Basic Research (under grant number 15-02-07740),  
and the UK Science and Technology Facilities Council (STFC) (under grant number 
ST/L005719/1). 
This material is based upon work supported by the U.S. Department of Energy, Office of Science, 
Office of Nuclear Physics under contract DE-AC05-06OR23177. 
\nocite{*}
\bibliographystyle{aipnum-cp}
\bibliography{gluex_meyer}
\end{document}